\pdfoutput=1
\documentclass[11pt,twoside,a4paper,cmspaper,final,collab]{cms-tdr}
\def\svnVersion{90b67ba}\def\svnDate{2021/09/28}\def\cmsCernNoTag{CERN-EP-2021-073}\def\cmsCernDate{\today}\def\cmsMessage{Published in the Journal of High Energy Physics as \href{http://dx.doi.org/10.1007/JHEP10(2021)174}{\doi{10.1007/JHEP10(2021)174}.}}
\begin{document}\cmsNoteHeader{SMP-19-013}

\newcommand{\fb}{\ensuremath{\,\text{fb}}\xspace}
\newcommand{\m}{\ensuremath{\,\text{m}}\xspace}
\newcommand{\T}{\ensuremath{\,\text{T}}\xspace}
\newcommand{\kHz}{\ensuremath{\,\text{kHz}}\xspace}
\newcommand{\MADSPIN}{{\textsc{MadSpin}}\xspace}
\newcommand{\PDFscale}{{\ensuremath{\,\text{(PDF + scale)}}\xspace}}
\newcommand{\MCstat}{{\ensuremath{\,\text{(MC stat)}}\xspace}}
\newcommand{\cmsTable}[1]{\resizebox{\textwidth}{!}{#1}}

\cmsNoteHeader{SMP-19-013}

\title{Measurements of the \texorpdfstring{$\Pp\Pp\to\PWpm\PGg\PGg$}{pp --> W gamma gamma} and \texorpdfstring{$\Pp\Pp\to\PZ\PGg\PGg$}{pp --> Z gamma gamma} cross sections at \texorpdfstring{$\sqrt s = 13\TeV$}{sqrt(s) = 13 TeV} and limits on anomalous quartic gauge couplings}

\date{\today}

\abstract{
The cross section for $\PW$ or $\PZ$ boson production in association with two photons is measured in proton-proton collisions at a centre-of-mass energy of 13\TeV. The data set corresponds to an integrated luminosity of 137\fbinv collected by the CMS experiment at the LHC. The $\PW\to\ell\PGn$ and $\PZ\to\ell\ell$ decay modes (where $\ell=\Pe,\PGm$) are used to extract the $\PW\PGg\PGg$ and $\PZ\PGg\PGg$ cross sections in a phase space defined by electron (muon) with transverse momentum larger than 30\GeV and photon transverse momentum larger than 20\GeV. All leptons and photons are required to have absolute pseudorapidity smaller than 2.5. The measured cross sections in this phase space are $\sigma(\PW\PGg\PGg)=13.6^{+1.9}_{-1.9}\stat^{+4.0}_{-4.0}\syst\pm 0.08\PDFscale\fb$ and $\sigma(\PZ\PGg\PGg)=5.41^{+0.58}_{-0.55}\stat^{+0.64}_{-0.70}\syst\pm 0.06\PDFscale\fb$. Limits on anomalous quartic gauge couplings are set in the framework of an effective field theory with dimension-8 operators.
}

\hypersetup{%
pdfauthor={CMS Collaboration},%
pdftitle={Measurements of the pp to W gamma gamma and pp to Z gamma gamma cross sections at  sqrt(s) =  13 TeV and limits on anomalous quartic gauge couplings},%
pdfsubject={CMS},%
pdfkeywords={CMS, standard model physics, multi boson, anomalous couplings}}

\maketitle

\section{Introduction}

The measurement of the associated production of a vector boson $\PV$ $(=\PW,\PZ)$ and two photons in proton-proton ($\Pp\Pp$) collisions is a powerful test of the standard model (SM). The nonabelian nature of the electroweak interaction predicts the presence of self-interacting vector boson vertices. The strength of the interaction is set by the values of triple and quartic gauge couplings predicted by the SM. The measurement of possible deviations from the theoretical predictions could provide indirect evidence of new particles or new interactions. Discrepancies at high photon momentum, where new physics might give a measurable deviation from the SM cross section, would produce evidence for the possible existence of anomalous quartic gauge couplings (aQGCs). A parametrisation of predictions involving anomalous couplings, independent of any specific new physics model, can be calculated in an effective field theory (EFT) framework~\cite{degrande201321}. Triboson production is also an important background for several SM and beyond the SM processes, such as the Higgs boson production in association with vector bosons (with $\PH\to\PGg\PGg$). Thus, studies of the self-couplings of the electroweak gauge bosons provide an excellent opportunity for a deeper understanding of electroweak interactions.

\begin{figure}[htb]
\centering
\includegraphics[width=0.40\textwidth]{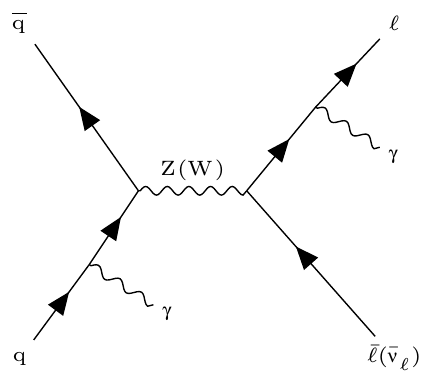}
\includegraphics[width=0.40\textwidth]{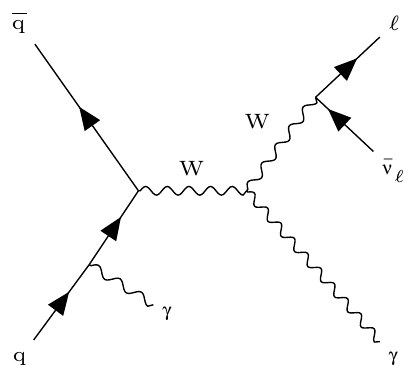}
\includegraphics[width=0.40\textwidth]{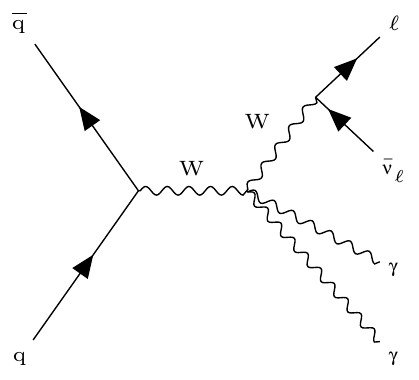}
\includegraphics[width=0.40\textwidth]{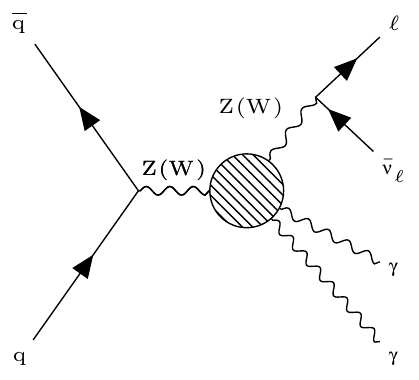}
\caption{Representative Feynman diagrams for the $\PV\PGg\PGg$ production in the SM (top left, top right and bottom left) and beyond the SM (bottom right).}
\label{f_vgg_feynman}
\end{figure}

Some of the elementary processes resulting in the production of a massive vector boson in association with two photons at the CERN LHC are presented in the leading-order (LO) Feynman diagrams of Fig.~\ref{f_vgg_feynman}. Topologies with final states that originate from events where the $\PV$ boson is produced in the hard interaction between the two partons, and the photons come from either initial or final state radiation processes or from quartic gauge vertices together with the $\PV$ boson, are studied, and are referred to as $\PV\PGg\PGg$ in the text.

Previous measurements of the $\PV\PGg\PGg$ production cross sections have been performed by the ATLAS and CMS Collaborations at the CERN LHC in $\Pp\Pp$ collisions at a centre-of-mass energy of $\sqrt s = 8\TeV$~\citep{atlas1,Aad:2016sau,cms1}. Limits on the presence of aQGCs were also reported in these papers.

In this paper, the first measurements of the $\Pp\Pp\to\PW\PGg\PGg$ and $\Pp\Pp\to\PZ\PGg\PGg$ cross sections at $\sqrt s = 13\TeV$ are presented using data collected between 2016 and 2018 by the CMS experiment, corresponding to an integrated luminosity of 137\fbinv. For these measurements, only direct decays into electrons or muons are considered. Measurements are compared with the latest available calculations at next-to-LO (NLO) in perturbative quantum chromodynamics (QCD)~\citep{qcd1,qcd2,qcd3}. Limits on the aQGCs are presented in the framework of an electroweak EFT with dimension-8 operators.

\section{The CMS detector}

The central feature of the CMS apparatus is a superconducting solenoid of 6\m internal diameter, providing a magnetic field of 3.8\T. Within the solenoid volume are a silicon pixel and strip tracker, a lead tungstate crystal electromagnetic calorimeter (ECAL), and a brass and scintillator hadron calorimeter (HCAL), each composed of a barrel and two endcap sections. Forward calorimeters extend the pseudorapidity $\eta$ coverage provided by the barrel and endcap detectors. Muons are detected in gas-ionisation chambers embedded in the steel flux-return yoke outside the solenoid.

Events of interest are selected using a two-tiered trigger system~\cite{Khachatryan:2016bia}. The first level, composed of custom hardware processors, uses information from the calorimeters and muon detectors to select events at a rate of around 100\kHz with a latency of about 4\mus. The second level, known as the high-level trigger, consists of a farm of processors running a version of the full event reconstruction software optimised for fast processing that reduces the event rate to around 1\kHz before data storage.

A more detailed description of the CMS detector, together with a definition of the coordinate system used and the relevant kinematic variables, is reported in Ref.~\cite{Chatrchyan:2008zzk}.

\section{Event simulation}

The associated production of a $\PW$ $(\PZ)$ boson and at least two photons is searched for in events with one lepton (two opposite-sign, same-flavour leptons) and two photons. Only electron and muon decay channels are used, while the $\PGt$ decays are treated as a background. The $\PV\PGg\PGg$ signal samples are generated at NLO with \MGvATNLO~\cite{MadGraph} (2.2.2 for 2016 samples, 2.2.6 for 2017 and 2018 samples) with no additional jets in the matrix element calculation. The NLO NNPDF 3.0 set~\cite{Ball2015} (for 2016 samples) and the next-to-NLO NNPDF 3.1 set~\cite{Ball:2017nwa} (for 2017 and 2018 samples) are used as parton distribution function (PDF) sets.

The main background contribution comes from the misidentification of jets as photons, which is estimated in single-photon control regions. Thus, various background samples involving a single photon are needed, including $\PV(\PV)\PGg$ samples. Other single-photon and diphoton processes (such as $\PQt\PAQt$ produced in association with one or two photons or a photon and a jet) contribute as backgrounds and are estimated using Monte Carlo (MC) simulations. The background contribution from the associated production of a $\PW(\PZ)$ boson and a Higgs boson decaying to two photons is negligible, and is not considered.

{\tolerance=1600
The $\PV\PGg$, the $\PQt\PGg$ and the $\PQt\PAQt\PGg$ samples are generated at NLO with \MGvATNLO with up to one additional jet in the matrix element calculation. The $\PV\PV\PGg$ and the $\PQt\PAQt\PGg\PGg$ samples are generated at NLO with \MGvATNLO with no additional jets in the matrix element calculation. The $\PGg$ plus jets samples are generated at LO with \MGvATNLO. The same PDF sets as for the signal samples are used. Alternative $\PV\PGg$ samples, which are generated with \SHERPA v2.2.6~\cite{Gleisberg_2004,Bothmann:2019yzt} at NLO precision with up to two additional jets and at LO precision for the three-jet computation using the NNPDF 3.1 set, are used for consistency checks and systematic uncertainties evaluations.
\par}

The \PYTHIA v.8.226 (v.8.230) package version is used for hadronisation with the CUETP8M1 tune~\cite{khachatryan:2015pea} (CP5 tune~\cite{sirunyan:2019dfx}) for the 2016 (2017 and 2018) samples.

Photons can be present also in other processes because of the hadronisation phase of the generation performed with \PYTHIA (v.8.2) even if not explicitly produced at matrix element level. To avoid possible double counting effects in the event selection, a procedure for the removal of the overlapping phase space region between inclusive and exclusive samples is implemented. Photons are selected at the generator level following a selection as close as possible to the one performed at reconstruction level. The total number of selected photons at the generator level is then used to remove the overlapping phase space between different samples. For single-photon processes (such as $\PW\PGg$ or $\PZ\PGg$), the event is discarded if the total number of selected photons at the generator level is different from one. The event is discarded from the diphoton processes if it has less than two photons selected at the generator level.

Predictions for aQGC signals are obtained by including a set of weights, corresponding to the presence of the anomalous couplings, to the $\PV\PGg\PGg$ reference samples simulated with \MGvATNLO. For this purpose, an aQGC model~\cite{degrande:2011ua} is used.

Additional $\Pp\Pp$ interactions in the same or adjacent bunch crossings (known as pileup) is included by adding simulated minimum bias events to the hard scattering. The events in the MC simulations are weighted so the distribution of the number of pileup interactions matches the one measured in data. The interaction of the particles with the CMS detector is simulated with \GEANTfour~\cite{agostinelli2003250}.

\section{Event selection}

Events for the $\PV\PGg\PGg$ analysis are selected using isolated single-lepton trigger requirements~\cite{Khachatryan:2016bia}. Single-electron trigger algorithms have a transverse-momentum \pt threshold of 27\GeV (for the 2016 data-taking period) and 32\GeV (for the 2017 and 2018 data-taking periods); single-muon trigger algorithms require \pt above 24\GeV for all three years.

All measured particles are reconstructed using the particle-flow (PF) algorithm~\cite{cms-prf-14-001}; this algorithm reconstructs and identifies each individual particle in an event with an optimised combination of information from the various elements of the CMS detector. The reconstructed vertex with the largest value of the sum of the $\pt^2$ of the physics objects is the primary $\Pp\Pp$ interaction vertex. The photon energies are obtained from the ECAL measurement. The electron energies are determined from a combination of the electron momentum at the primary interaction vertex as determined by the tracker, the energy of the corresponding ECAL cluster, and the energy sum of all bremsstrahlung photons spatially compatible with the ones originating from the electron track. The muon energies are obtained from the curvature of the muon tracks.

Electrons candidates are required to have $\pt>15\GeV$ in the pseudorapidity ranges that exclude the barrel-endcap transition region, $\abs{\eta}<1.44$ and $1.57<\abs{\eta}<2.50$. A variety of criteria is used to separate genuine electrons from misidentified ones.  A tight identification is used to select prompt electrons (produced at the primary vertex) and isolated electrons in the final state~\cite{Sirunyan:2020ycc}.
Background contributions from misidentified jets or electrons inside a jet are rejected applying electron isolation criteria, which exploit the PF-based event reconstruction. The electron isolation variables are obtained by summing the \pt of charged hadrons compatible with the primary vertex $I_{\text{chg}}$, of neutral hadrons $I_{\text{neu}}$, and of photons $I_{\text{pho}}$ inside a cone of radius $\Delta R=\sqrt{\smash[b]{(\Delta\eta)^2+(\Delta\phi)^2}}=0.3$ around the electron direction, where $\phi$ is the azimuthal angle in radians. Additional photons and neutral hadronic contributions to the isolation variable, coming from pileup, are subtracted using the jet area approach~\cite{cacciari:2007fd}.

Muons candidates are required to have $\pt>15\GeV$ in the pseudorapidity range $\abs{\eta}<2.4$. Muon identification criteria are based on the fit quality for tracks measured in the tracker and muon detectors. A tight muon identification is used to reconstruct muons in the final state~\cite{Sirunyan:2018fpa}.
To distinguish between prompt muons and those from hadron decays within jets, muons are required to be isolated with respect to all nearby PF reconstructed particles.
 For the computation of the PF isolation, the $I_{\text{chg}}$, $I_{\text{neu}}$ and $I_{\text{pho}}$ components are summed in a cone of $\Delta R = 0.4$ around the muon direction. The corrected energy sum is obtained by subtracting the pileup contribution to $I_{\text{neu}}$ and $I_{\text{pho}}$, which is estimated as half of the corresponding charged hadronic component.

Photons are selected with $\pt>20\GeV$ in the pseudorapidity range $\abs{\eta}<1.44$ and $1.57<\abs{\eta}<2.5$. Photon identification is based on the sequential application of several selections. A medium photon identification is used to reconstruct prompt photons (\ie not from hadron decays) in the final state~\cite{Sirunyan:2020ycc}. The average efficiency for this selection is 80\%. Photons selected in this analysis are required to have a narrow transverse shape of the electromagnetic shower, a minimal energy deposit in the HCAL, and to be isolated with respect to other particles. The same isolation variable previously described for the electron selection is used.

The reconstruction, identification, and isolation efficiencies of leptons and photons and the trigger efficiencies of leptons are measured with the ``tag-and-probe'' technique \cite{chatrchyan2011}, as a function of particle $\eta$ and \pt in both data and simulation. A sample of events containing a $\PZ$ boson decaying into $\Pep\Pem$ or $\PGmp\PGmm$ is used for these measurements. The photon efficiencies are derived using a sample of electrons from $\PZ$ decays with no requirement on the track and charge of the candidate. These efficiencies are used to correct for the differences between data and simulation.

An event is categorised as a $\PW$ decaying to leptons if exactly one electron (muon) with $\pt>35\,(30)\GeV$ is selected. The selected lepton must match the one that triggered the event and must be associated with the primary vertex of the collision. If the event contains any additional different-flavour leptons or opposite-sign same-flavour leptons, it is excluded from the $\PW$ boson candidate sample, but is further checked for the presence of a $\PZ$ boson candidate.

An event is categorised as a $\PZ$ boson candidate decaying to leptons if two opposite-sign leptons of the same flavour are selected. The leading \pt electron (muon) is required to have $\pt>35\,(30)\GeV$, and the subleading electron or muon is required to have $\pt>15\GeV$. Only the leading lepton is required to match the one that triggered the event, although both are required to be associated with the primary vertex of the collision. The invariant mass of the dilepton system is required to be $m_{\ell\ell}>55\GeV$.

Events are selected if they have a single $\PW$ or $\PZ$ boson candidate and at least two photons. All reconstructed photons must be separated from each other and from each reconstructed lepton by $\Delta R>0.4$. Photons are discarded if $\abs{m_{\Pe,\PGg}-m_{\PZ}}<5\GeV$ (where $m_{\Pe,\PGg}$ is the invariant mass of an electron and the leading or subleading photon and $m_{\PZ}$ is the $\PZ$ boson invariant mass) or if $\abs{m_{\Pe\PGg\PGg}-m_{\PZ}}<5\GeV$ (where $m_{\Pe\PGg\PGg}$ is the invariant mass of an electron and the two photons). In this way, photons likely coming from final-state bremsstrahlung radiation are removed and, therefore, the contribution from electrons misidentified as photons is reduced as well.

\section{Background estimation}\label{s_background}

The backgrounds in both the $\PW\PGg\PGg$ and $\PZ\PGg\PGg$ signal regions are categorised as events with a genuine photon or with another object misidentified as a photon. 
The largest contribution in both channels comes from the misidentification of jets as photons. 
Another important source of background originates from electrons that are reconstructed as photons because the deposit in the calorimeter is not associated with a track in the tracker. 
This contribution is particularly relevant in the $\PW\PGg\PGg$ electron channel, and is dominated by the $\PZ\PGg$ electron channel.
Both of these background processes are estimated by exploiting a control sample in data. 
The remaining minor contributions from processes that have at least one genuine photon ($\PQt\PGg$, $\PQt\PAQt\PGg$, $\PQt\PAQt\PGg\PGg$, $\PV\PGg$ and $\PV\PV\PGg$) are estimated using MC simulations and referred to as ``others''.

The background from events containing nonprompt photons is estimated following a method similar to the ones described in Refs.~\cite{cms1,Aad:2016sau}. 
A $\PW$ or a $\PZ$ boson is selected together with one photon that passes the standard selection except for the isolation requirement both in data and simulation. 
Events are categorised as ``tight'' or ``loose'' if the photon candidate passes or fails the isolation requirement.
The probabilities for photon $\epsilon$ and for a jet $f$ to be isolated are computed as
\begin{linenomath}
\begin{equation*}
\epsilon=\frac{N_{\PGg,\,\mathrm{MC}}^\mathrm{T}}{N_{\PGg,\,\mathrm{MC}}^\mathrm{T}+N_{\PGg,\,\mathrm{MC}}^\mathrm{L}}\text{ and }f=\frac{N_{\PGg,\,\text{data}}^\mathrm{T}}{\mathrm{N}_{\PGg,\,\text{data}}^\mathrm{T}+N_{\PGg,\,\text{data}}^\mathrm{L}},
\end{equation*}
\end{linenomath}
where $N_{\PGg,\,\mathrm{MC}}^\mathrm{T}$ ($N_{\PGg,\,\mathrm{MC}}^\mathrm{L}$) is the number of simulated events with a tight (loose) photon while $N_{\PGg,\,\text{data}}^\mathrm{T}$ ($N_{\PGg,\,\text{data}}^\mathrm{L}$) is the number of events with a tight (loose) photon candidate in data after the subtraction of the prompt photon contribution from simulation.
These probabilities are calculated separately for photons in the ECAL barrel and endcap regions as a function of the photon \pt.
The jet-photon misidentification background in the diphoton phase space is then estimated by solving the system
\begin{linenomath}
\begin{equation*}
\begin{pmatrix}
N_{\mathrm{TT}}\\
N_{\mathrm{TL}}\\
N_{\mathrm{LT}}\\
N_{\mathrm{LL}}
\end{pmatrix}
=
\begin{pmatrix}
\epsilon_1\epsilon_2&\epsilon_1f_2&f_1\epsilon_2&f_1f_2\\
\epsilon_1(1-\epsilon_2)&\epsilon_1(1-f_2)&f_1(1-\epsilon_2)&f_1(1-f_2)\\
(1-\epsilon_1)\epsilon_2&(1-\epsilon_1)f_2 & (1-f_1)\epsilon_2&(1-f_1)f_2\\
(1-\epsilon_1)(1-\epsilon_2)&(1-\epsilon_1)(1-f_2)&(1-f_1)(1-\epsilon_2)&(1-f_1)(1-f_2)\\
\end{pmatrix}
\begin{pmatrix}
\alpha_{\PGg\PGg}\\
\alpha_{\PGg\mathrm{j}}\\
\alpha_{\mathrm{j}\PGg}\\
\alpha_{\mathrm{jj}}
\end{pmatrix},
\end{equation*}
\end{linenomath}
where the indices of the $\epsilon$ and $f$ coefficients refer to the leading and the subleading photon. 
The $N_\mathrm{XY}$ vector contains the number of events where two ($\mathrm{TT}$), one ($\mathrm{TL}$ and $\mathrm{LT}$) or zero ($\mathrm{LL}$) photon candidates pass the isolation requirement.
The $\alpha_\mathrm{AB}$ vector contains the number of signal ($\PGg\PGg$) and background ($\PGg\mathrm{j}$, $\mathrm{j}\PGg$ and $\mathrm{jj}$) events.
This method is validated in a control region enriched in the jet-photon misidentification background where both photons fail the isolation selection. 
By solving the matrix for each combination of the single photons \pt and $\eta$ and by applying it to the number of events in the double-photon phase space in each diphoton \pt bin one determines the number of background events in the signal region as
\begin{linenomath}
\begin{equation*}
N_{\mathrm{TT}}^{\mathrm{j}\to\PGg}=\sideset{}{_{(\pt, \eta)_{\PGg1}\,(\pt, \eta)_{\PGg2}} }\sum(\epsilon_1 f_2 \alpha_{\PGg\mathrm{j}} + f_1 \epsilon_2 \alpha_{\mathrm{j}\PGg} + f_1 f_2 \alpha_{\mathrm{jj}}).
\end{equation*}
\end{linenomath}
Because of the large contamination from $\PZ\PGg\to\Pe\Pe\PGg$ events where an electron from the $\PZ$ boson decay is misclassified as a photon, the jet misidentified as a photon contribution for the $\PW\PGg\PGg$ electron channel is estimated from the $\epsilon$ and $f$ coefficients that are evaluated from the single-photon $\PW\PGg$ muon sample. For the same reason, the $\PZ\PGg\to\Pe\Pe\PGg$ background contribution has been subtracted from data in the $\PW\PGg\PGg$ electron channel before computing the number of background events in the signal region.

To estimate the contribution where an electron is misclassified as a photon, a correction factor is computed from $\PZ\PGg$ events and is then applied to the simulation. The invariant mass of an electron and a photon is reconstructed in data and MC simulation while removing the requirement $\abs{m_{\Pe,\,\text{lead}\,\PGg}-m_{\PZ}}<5\GeV$. This mass distribution is fitted with the sum of a signal template, derived from MC simulation, and a background function, which has an exponential decay distribution at high mass (above the $\PZ$ peak) and a turn-on (linked to the electron and photon \pt thresholds) described by an error function at low mass. A correction factor, obtained in intervals of the photon \pt and $\eta$, is then computed as
\begin{linenomath}
\begin{equation}
\mathcal{F}(\pt,\eta)=\frac{N_{\text{inv}}^\text{data}/N_{\PZ}^\text{data}}{N_{\text{inv}}^\mathrm{MC}/N_{\PZ}^\mathrm{MC}},
\label{eq:F_factors}
\end{equation}
\end{linenomath}
where $N_{\text{inv}}^\text{data}$ is the number of events in the electron-photon invariant mass peak obtained by integration of the fitted signal shape and $N_{\PZ}^\text{data}$ is the number of events for the $\PZ\to\Pe\Pe$ invariant mass distribution obtained by integration of a fitted double-sided Crystal-Ball function~\cite{Oreglia:1980cs} in the data. The same procedure is used to calculate the number of events in MC simulation. By fitting all the distributions in the different $(\pt,\eta)$ bins, a set of correction factors is computed.
All the MC simulations are then corrected for these factors on an event-by-event basis whenever a reconstructed photon matches a generator-level electron.
The event-by-event correction factors are on average about 20\%.

\begin{figure}[htb]
\centering
\includegraphics[width=0.48\textwidth]{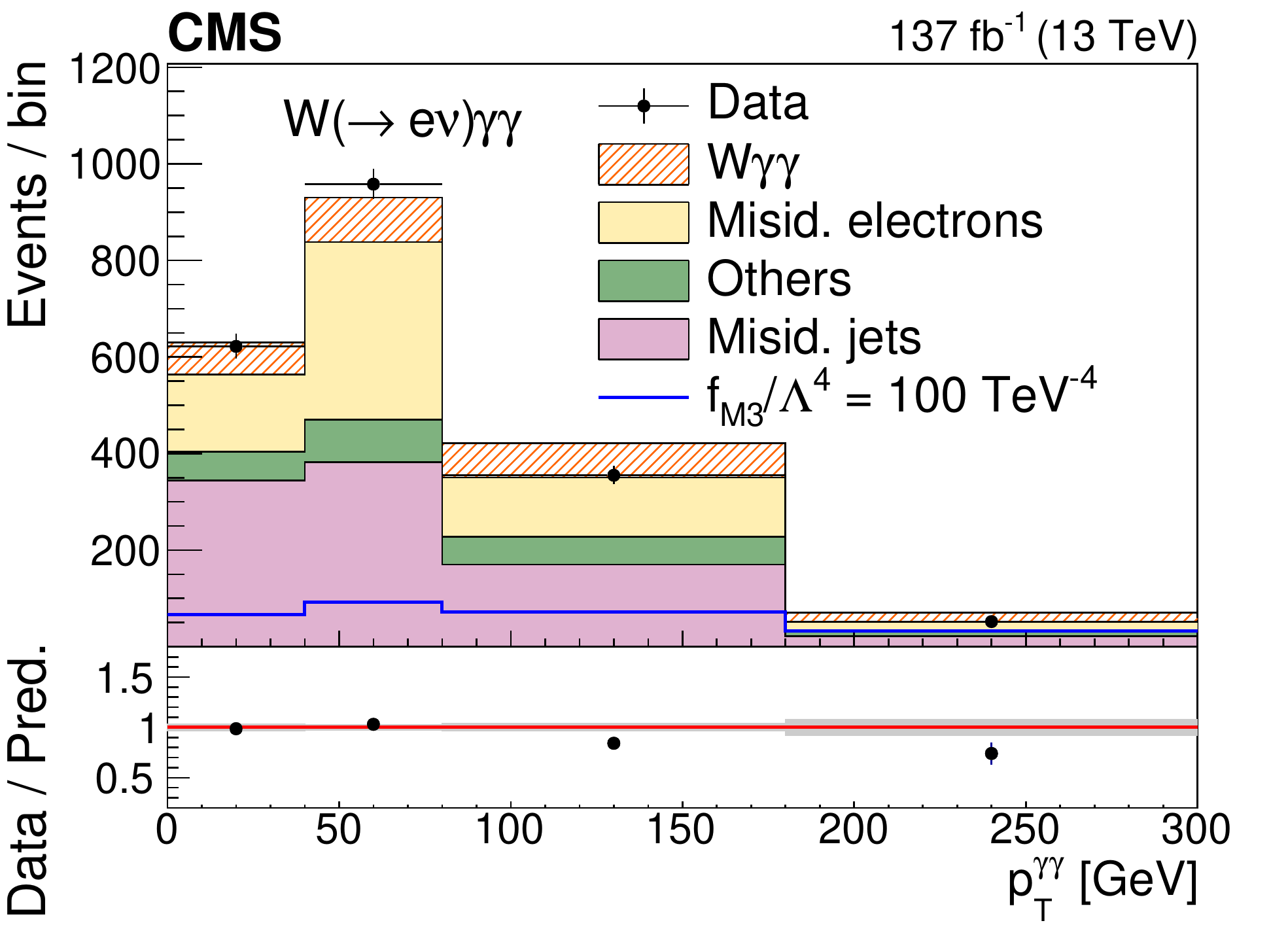}
\includegraphics[width=0.48\textwidth]{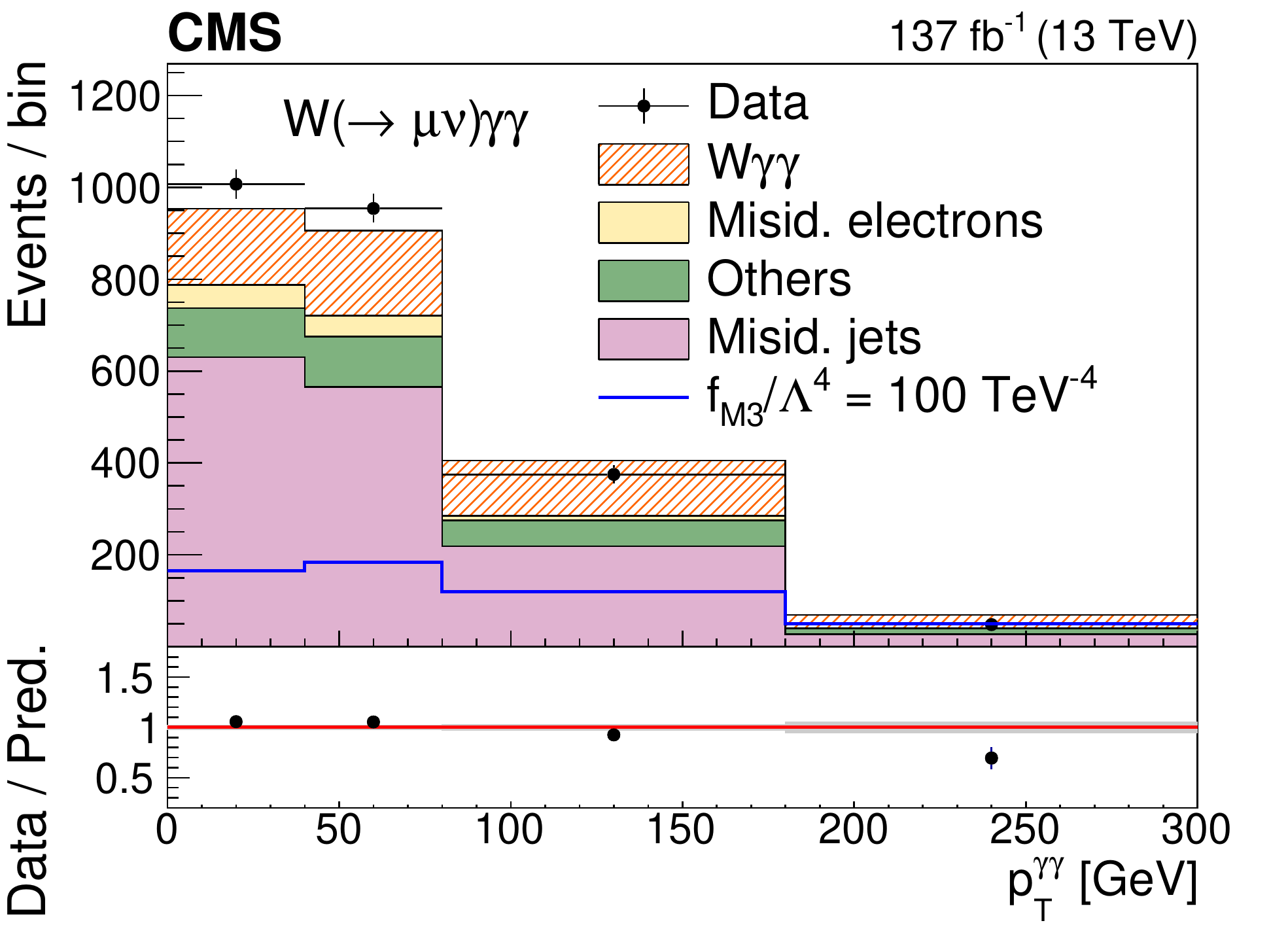}
\includegraphics[width=0.48\textwidth]{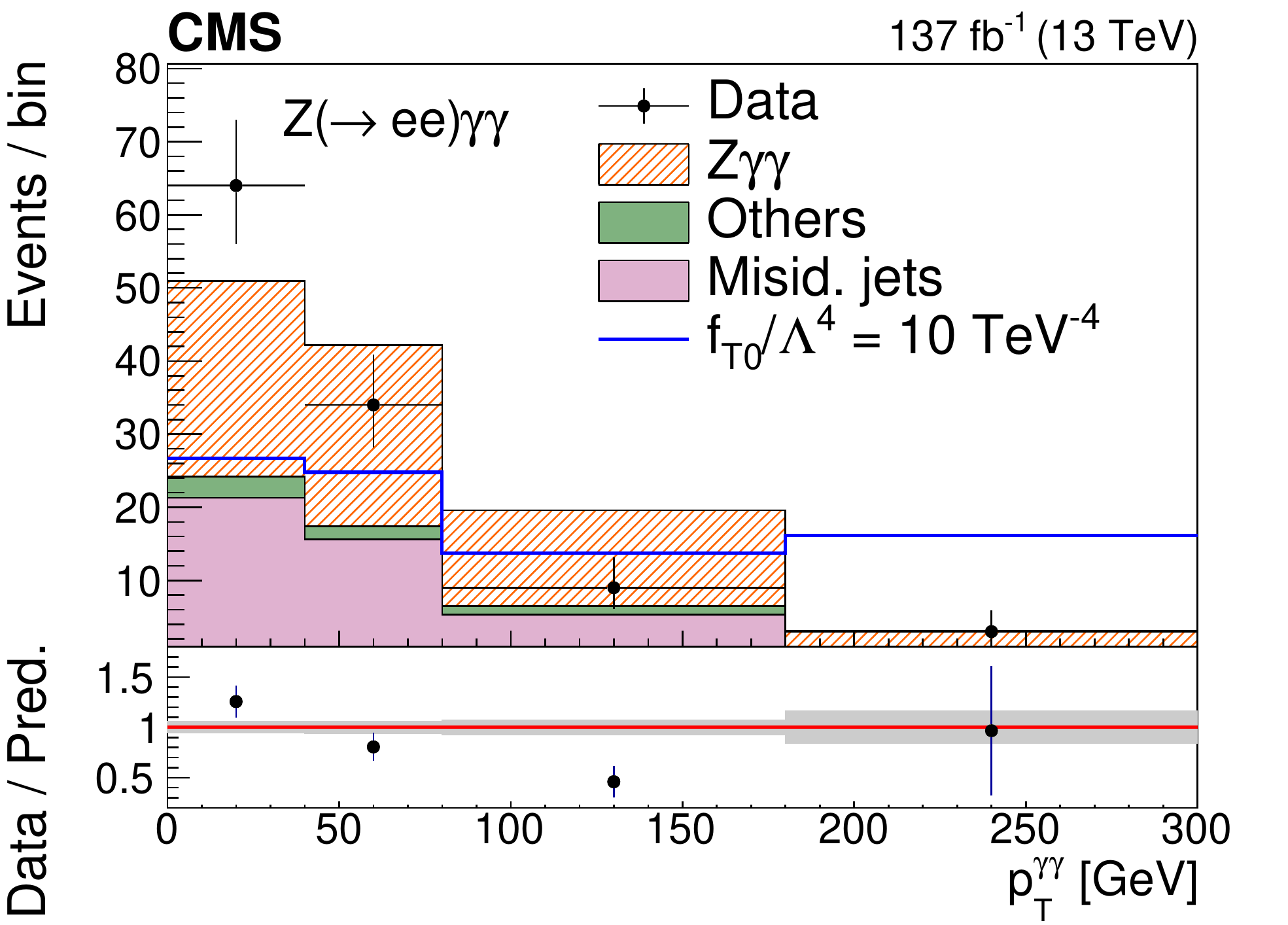}
\includegraphics[width=0.48\textwidth]{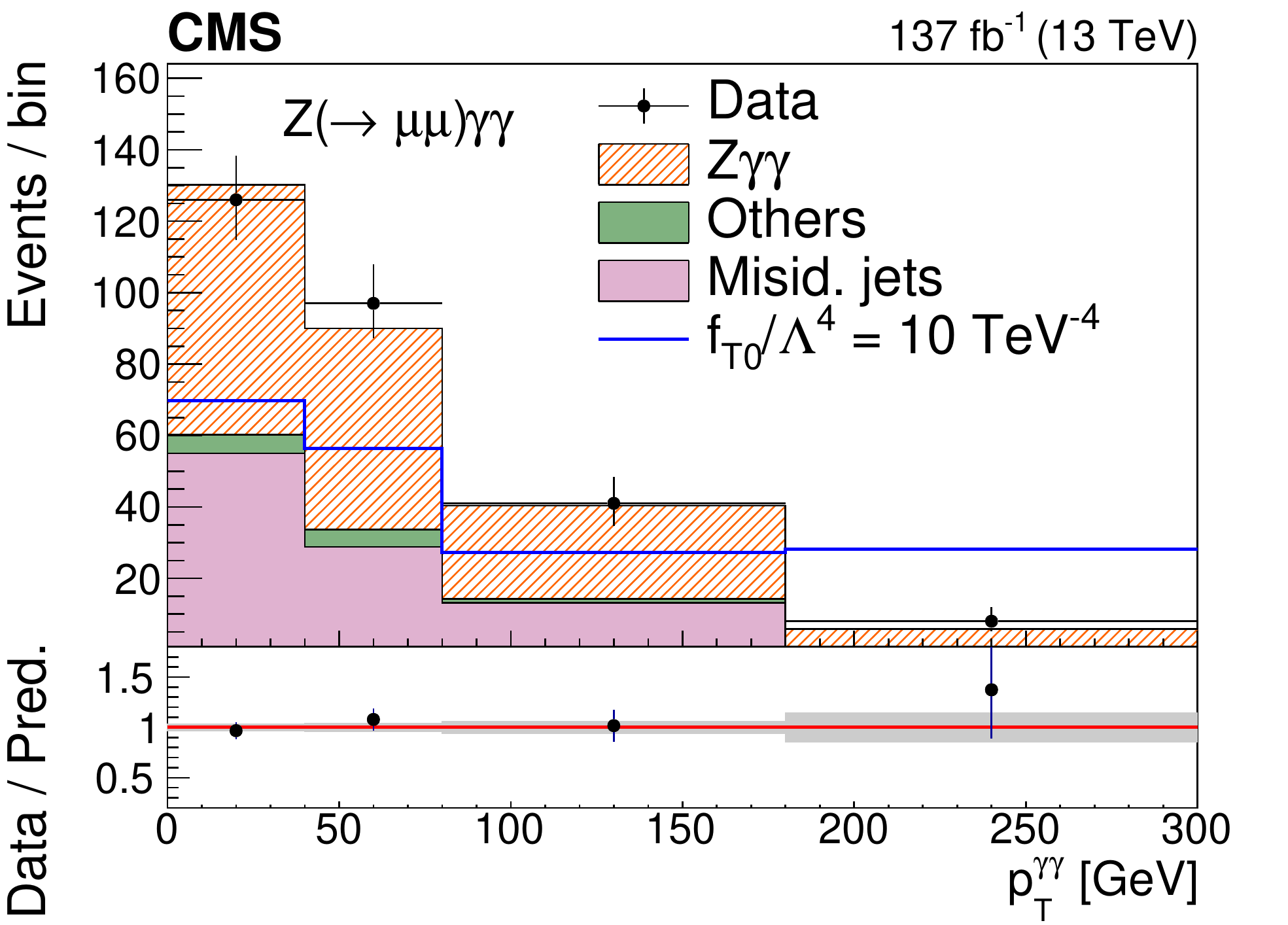}
\vspace*{\baselineskip}
\caption{Distribution of the transverse momentum of the diphoton system for the $\PW\PGg\PGg$ electron (upper left) and muon (upper right) channels and for the $\PZ\PGg\PGg$ electron (lower left) and muon (lower right) channels.
The predicted yields are shown with their pre--fit normalisations.
The black points represent the data with error bars showing the statistical uncertainties.
The hatched histogram shows the expected signal contribution.
The background estimate for electron (jet) misidentified as photons, obtained from control samples in data, is shown in yellow (purple). The remaining background, derived from MC simulation, is shown in green.
In the ratio plots, the grey hashed area is the statistical uncertainty on the sum of signal and backgrounds, while the uncertainty in the black dots is the statistical uncertainty of the data. In blue, the expected distribution for an example value of the anomalous coupling parameters $f_{\mathrm{M}3}/\Lambda^4$ and $f_{\mathrm{T}0}/\Lambda^4$ is also shown (see Section~\ref{sec:limits} for the details).}
\label{f_vgg_diphopt}
\end{figure}

\begin{figure}[htb]
\centering
\includegraphics[width=0.48\textwidth]{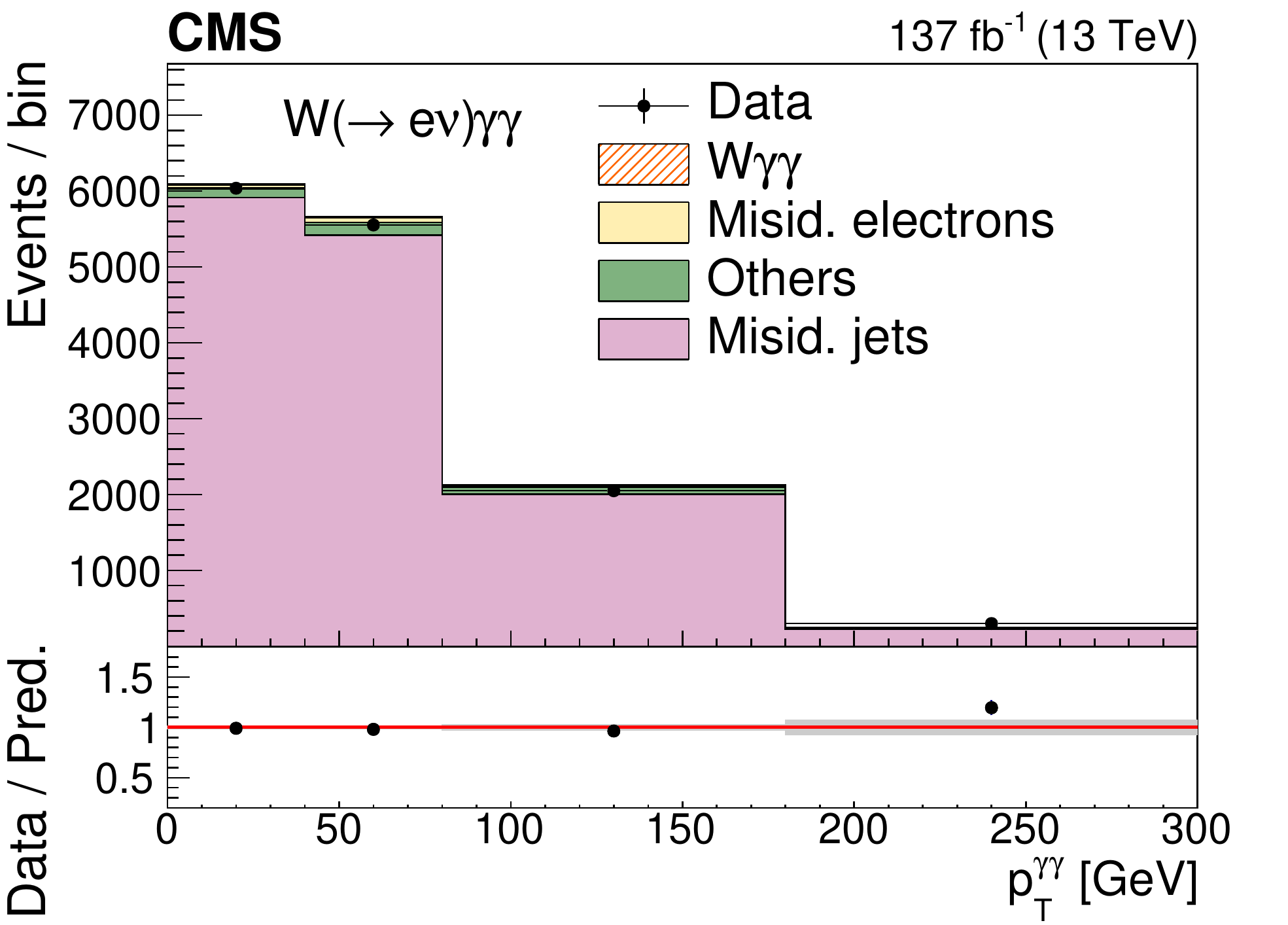}
\includegraphics[width=0.48\textwidth]{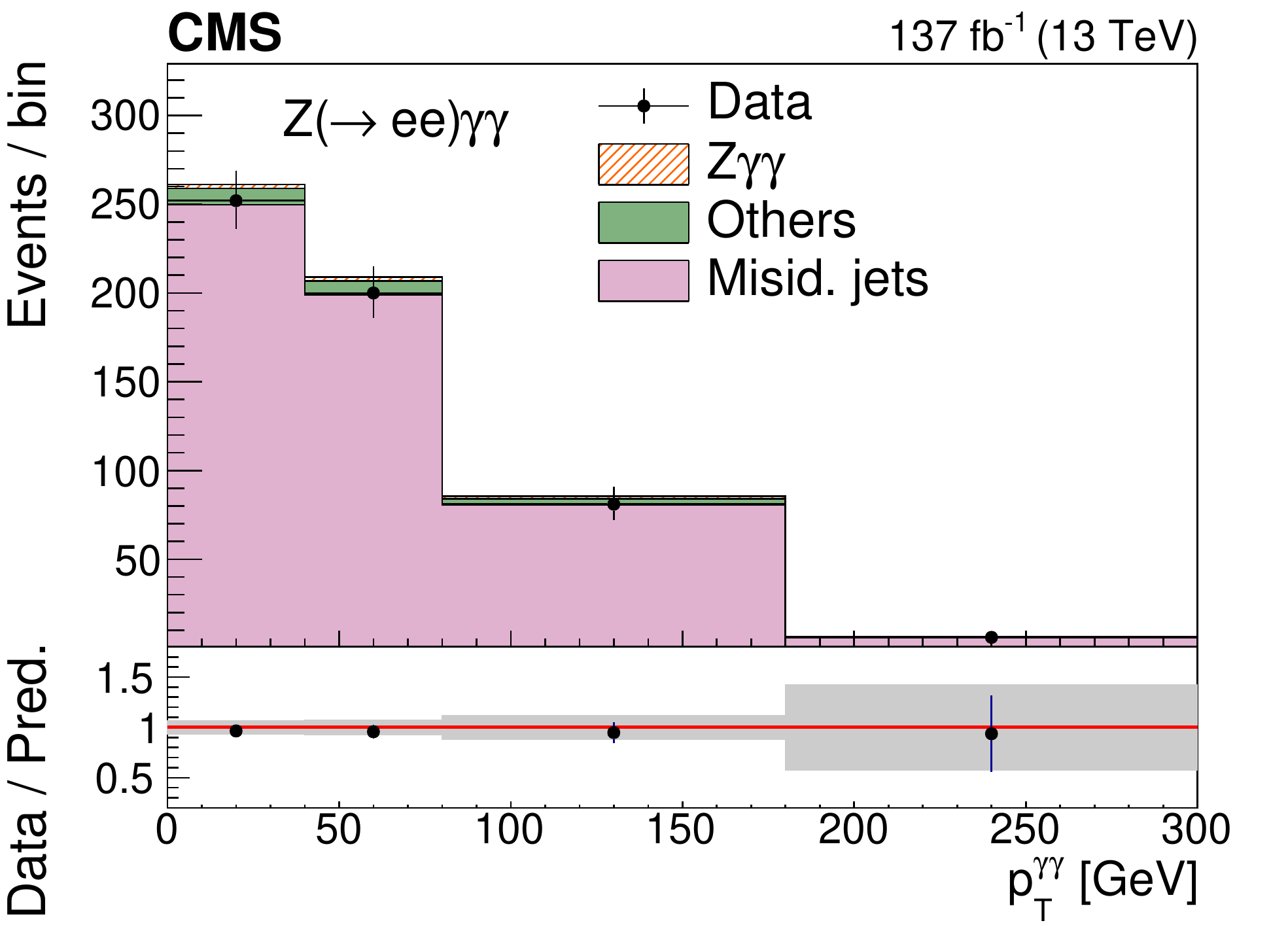}
\vspace*{\baselineskip}
\caption{Distribution of the transverse momentum of the diphoton system, obtained in the control region enriched in jets misidentified as photons, for the $\PW\PGg\PGg$ and for the $\PZ\PGg\PGg$ electron channels.
The black points represent the data with error bars showing the statistical uncertainties.
The hatched histogram shows the expected negligible signal contribution.
The background estimate for electron (jet) misidentified as photons, obtained from control samples in data, is shown in yellow (purple). The remaining backgrounds, derived from MC simulation, are shown in green.
In the ratio plots, the grey hashed area is the statistical uncertainty on the sum of signal and backgrounds, while the uncertainty in the black dots is the statistical uncertainty of the data.}
\label{f_vgg_validation}
\end{figure}

The pre--fit (\ie before the fitting procedure described in Section~\ref{sec:crosssec}) diphoton \pt distributions for the $\PW\PGg\PGg$ and $\PZ\PGg\PGg$ analyses, separated in the electron and muon channels, are shown in Fig.~\ref{f_vgg_diphopt}.
The same distributions obtained in the control region enriched in jets misidentified as photons are shown in Fig.~\ref{f_vgg_validation} for the $\PW\PGg\PGg$ and $\PZ\PGg\PGg$ electron channels.
The data and prediction agree thus validating the jet--photon misidentification background estimation procedure.
A similar level of agreement is observed in the muon channel.

\section{Systematic uncertainties}

Systematic effects can affect the rates and distributions of both data and simulation. To estimate these uncertainties, the full analysis procedure is repeated by varying each quantity by plus or minus its standard deviation uncertainty. In this procedure, correlations between the systematic uncertainties are included where appropriate. 

The dominant systematic uncertainties come from the estimation of the backgrounds. To determine the systematic uncertainty coming from the jet-photon misidentification background, the same strategy is applied to a QCD control sample that is obtained using the $\PW\PGg$ selection but inverting the isolation requirement on the leptons while keeping the photon selection identical to the one for the signal region. This sample is used to obtain an alternative estimate of the jet-photon misidentification background contribution in the $\PW$ channel. For the $\PZ$ channel, the QCD control sample resulting from the $\PZ\PGg$ selection with the inversion of the lepton isolation has insufficient events. Hence, a transfer factor from the $\PW\PGg$ selection is computed and applied for the determination of the alternative estimate of the jet-photon misidentification background contribution in the $\PZ$ channel. The systematic uncertainty is computed as half the difference in the distributions between the standard method or the one just described.

Another source of uncertainty in the jet-photon background is related to the modelling of the initial- and final-state radiation and of the energy spectra of the final state particles. An alternative $\PV\PGg$ MC simulation, obtained with \SHERPA, is used to evaluate this uncertainty.

The uncertainty in the correction factor related to the background of electrons misidentified as photons is determined by propagating the estimated uncertainty in the correction factor $\mathcal{F}$ of Eq.~\ref{eq:F_factors}. This has two components: a statistical one, that comes from the uncertainty in the fitting procedure; and a systematic one that is computed by taking half the difference between the $\mathcal{F}$ factors obtained by performing the fit of the signal component with a double-sided Crystal-Ball function and with the nominal method using an MC template.

The uncertainties in the lepton and photon reconstruction and selection efficiencies are included by computing the cross section with these efficiencies varied up and down by one standard deviation. The uncertainty related to these data-to-simulation corrections is estimated by including the uncertainty in the tag-and-probe method. Uncertainties in the trigger efficiencies are negligible.

The uncertainty in the value of the theoretically computed cross section is accounted for during the subtraction from data of the background processes estimated from MC.
Furthermore, the value of the expected cross section has an impact on the estimation of the jet-photon background because the contribution from prompt photons is subtracted from the distribution in data using the $\PW\PGg$ and $\PZ\PGg$ simulations. 
To estimate these contributions, the cross sections of the $\PW\PGg$, $\PZ\PGg$, and of the other minor backgrounds are varied independently. 
The uncertainty in the $\PZ\PGg$ cross section is estimated as half the difference between the next-to-NLO and the NLO values computed with MATRIX~\cite{Grazzini_2018}, and amounts to 2.5\%.
The same uncertainty is assumed for the $\PW\PGg$ cross section, and a value of 7.5\% is used for the other simulated backgrounds.

The total inelastic cross section is varied by 4.6\%~\cite{Sirunyan_2018inpp} to estimate the impact on the final result of the pileup reweighting procedure. The uncertainty because of the integrated luminosity measurement is equal to 2.5, 2.3, and 2.5\% for the 2016, 2017 and 2018 data taking periods, respectively~\cite{CMS-PAS-LUM-17-001,CMS-PAS-LUM-17-004,CMS-PAS-LUM-18-002,CMS:2021xjt}. Because of the uncorrelated time evolution of some systematic uncertainties, the total integrated luminosity has an uncertainty of 1.8\% and is applied to all the processes estimated with an MC simulation. The effect of the uncertainty in the integrated luminosity affects the estimation of the jet-photon misidentification background as well as the MC contributions in the diphoton distributions.

For the extraction of the results, each systematic uncertainty is represented by a nuisance parameter, which affects the shape and the normalisation of the distribution of the various background contributions. The variation of the nuisance parameter results in a continuous perturbation of the spectrum, following a Gaussian probability density function. The impact of each systematic uncertainty is obtained by freezing the set of associated nuisance parameters to their best-fit values and comparing the total uncertainty in the measured cross section with the result from the nominal fit~\cite{Prosper:1306523}.
The contributions of the different systematic uncertainties for both the $\PW\PGg\PGg$ and $\PZ\PGg\PGg$ processes are presented in Table~\ref{t_syst}.

\begin{table}
\centering
\topcaption{Summary of the systematic uncertainties (in percent) for the $\PW\PGg\PGg$ and $\PZ\PGg\PGg$ cross section measurements. The numbers indicate the impact of each systematic uncertainty in the value of the measured cross section in the corresponding channel. The systematic uncertainties in the jets misidentified as photons are added in quadrature in the table.}
\cmsTable{
\begin{tabular}{ccccccc}
Systematic source & $\Pe\PGne\PGg\PGg\,[\%]$ & $\PGm\PGnGm\PGg\PGg\,[\%]$ & $\ell\PGn\PGg\PGg\,[\%]$ & $\Pe\Pe\PGg\PGg\,[\%]$ & $\PGm\PGm\PGg\PGg\,[\%]$ & $\ell\ell\PGg\PGg\,[\%]$\\
\hline
Integrated luminosity & $<$1 & 2 & 2 & 3 & 1 & 3\\
Pile-up & 2 & $<$1 & $<$1 & 2 & $<$1 & 1\\
Electron efficiencies & 4 & \NA & $<$1 & 3 & \NA & 1\\
Muon efficiencies & 1 & $<$1 & $<$1 & 2 & $<$1 & 1\\
Photon efficiencies & 18 & 13 & 12 & 6 & 5 & 5\\
Jet-photon misid. & 25 & 22 & 21 & 6 & 5 & 6\\
Electron-photon misid. & 4 & $<$1 & $<$1 & \NA & \NA & \NA\\
$\PW\PGg$ theoretical cross section & 3 & 3 & 3 & $<$1 & $<$1 & $<$1\\
$\PZ\PGg$ theoretical cross section & 4 & $<$1 & $<$1 & 7 & 5 & 6\\
Other bkgs theoretical cross section & 5 & 2 & 2 & $<$1 & $<$1 & $<$1 \\
Simulated sample event count & 18 & 7 & 8 & 7 & 3 & 4\\
\end{tabular}
}
\label{t_syst}
\end{table}

\section{Cross section measurements}\label{sec:crosssec}

The cross sections for the $\PW\PGg\PGg$ and $\PZ\PGg\PGg$ processes are measured separately in the electron and muon channels using a sample of events corresponding to an integrated luminosity of 137\fbinv (LHC Run 2 data). The observed and predicted numbers of events are presented in Table~\ref{t_vgg_evt}.

\begin{table}
\topcaption{Summary of the pre--fit predicted and observed numbers of events for 137\fbinv for the $\PW\PGg\PGg$ (upper Table) and $\PZ\PGg\PGg$ (lower Table) selections in the electron and muon channels. The systematic uncertainties of the individual backgrounds and the total background are obtained by summing the contributions of different systematic uncertainties in quadrature. The statistical uncertainties are those related to the MC event samples and control region statistical uncertainties.}
\centering
\begin{tabular}{ccc}
Process & $\Pe\PGne\PGg\PGg$ & $\PGm\PGnGm\PGg\PGg$\\
\hline
Misid. jets & $918\pm 23\stat\pm 180\syst$ & $1441\pm 27\stat\pm 280\syst$\\
Misid. electrons & $669\pm 28\stat\pm 34\syst$ & $107\pm 9\stat\pm 7\syst$\\
Others & $217\pm 11\stat\pm 20\syst$ & $286\pm 11\stat\pm 25\syst$\\
Total backgrounds & $1804\pm 38\stat\pm 180\syst$ & $1834\pm 30\stat\pm 280\syst$\\
Expected signal & $248\pm 6\stat\pm 17\syst$ & $500\pm 8\stat\pm 33\syst$\\
Total prediction & $2052\pm 38\stat\pm 180\syst$ & $2334\pm 31\stat\pm 280\syst$\\
Data & 1987 & 2384\\
\hline
\hline
Process & $\Pe\Pe\PGg\PGg$ & $\PGm\PGm\PGg\PGg$\\
\hline
Misid. jets & $42\pm 4\stat\pm 9\syst$ & $98\pm 5\stat\pm 27\syst$\\
Others & $6\pm 1\stat\pm 1\syst$ & $11\pm 2\stat\pm 1\syst$\\
Total backgrounds & $48\pm 4\stat\pm 9\syst$ & $109\pm 6\stat\pm 27\syst$\\
Expected signal & $68\pm 2\stat\pm 5\syst$ & $157\pm 3\stat\pm 11\syst$\\
Total prediction & $116\pm 4\stat\pm 8\syst$ & $266\pm 6\stat\pm 23\syst$\\
Data & 110 & 272\\
\end{tabular}
\label{t_vgg_evt}
\end{table}

The measured yields in the electron and muon channels are extrapolated to a common fiducial phase space determined from simulated signal events at the generated particle level. Generated particles are considered stable if their mean decay length is larger than 1\cm. Generated leptons are required to have a $\pt>15\GeV$ and $\abs{\eta}<2.5$. The momenta of photons in a cone of $\Delta R=0.1$, the same cone size as the one applied to reconstructed data, are added to the charged lepton momentum to correct for final-state radiation. Generated photons are required to have $\pt>15\GeV$ and $\abs{\eta}<2.5$. Additionally, the candidate photons are required to have no selected leptons or photons in a cone of radius $\Delta R=0.4$ and no other stable particles, apart from photons and neutrinos, in a cone of radius $\Delta R=0.1$. Events are then selected in the $\PW\PGg\PGg$ channel by requiring exactly one electron (muon) with $\pt>30\GeV$ and at least two photons with $\pt>20\GeV$. Events are selected in the $\PZ\PGg\PGg$ channel by requiring two electrons (muons), at least one of them with $\pt>30\GeV$, and not less than two photons, each of them with $\pt>20\GeV$. Additionally, the invariant mass of the dilepton system is required to be $m_{\ell\ell}>55\GeV$.

The expected theoretical cross sections are predicted at NLO and their uncertainties come from the finite MC sample event count used to compute them, the PDF set, the factorisation and renormalisation scales. Statistical uncertainties are estimated to be of the order of 0.2\% in both the $\PW\PGg\PGg$ and $\PZ\PGg\PGg$ channels. Uncertainties related to the PDF set are estimated using a set of 100 replicas of the NNPDF~3.1 PDF set, following the Ref.~\cite{Ball:2017nwa} prescription, and are estimated to be of the order of 0.3\% in the $\Pe\PGn\PGg\PGg$ and $\PGm\PGn\PGg\PGg$ and of 0.8\% in the $\Pe\Pe\PGg\PGg$ and $\PGm\PGm\PGg\PGg$ channels. Uncertainties related to the renormalisation and factorisation scale choice are estimated by independently varying $\mu_\mathrm{R}$ and $\mu_\mathrm{F}$ by a factor of 0.5 and 2, with the condition that $1/2<\mu_\mathrm{R}/\mu_\mathrm{F}<2$. The uncertainties are defined as the maximal differences from the nominal values and are estimated to be of the order of 0.6 (0.5)\% in the $\Pe\PGn\PGg\PGg$ ($\PGm\PGn\PGg\PGg$) and of 0.6 (0.7)\% in the $\Pe\Pe\PGg\PGg$ ($\PGm\PGm\PGg\PGg$) channels. Uncertainties related to the value of the strong coupling are estimated to be of the order of 0.03 (0.02)\% in the $\Pe\PGn\PGg\PGg$ ($\PGm\PGn\PGg\PGg$) and of 0.4\% in the $\Pe\Pe\PGg\PGg$ and $\PGm\PGm\PGg\PGg$ channels.

Binned maximum likelihood fits to the diphoton \pt distributions in Fig.~\ref{f_vgg_diphopt} are performed to extract the signal strength $\mu$ and the significance of the results~\cite{CMS-NOTE-2011-005, Khachatryan:2014jba}. The results are obtained separately in the electron, muon and lepton channels. The systematic uncertainties and the statistical uncertainty in the MC predictions are treated as nuisance parameters in the fits and profiled. The high \pt bins in the distributions are the more relevant ones for the determination of the limits.

{\tolerance=800
The measured cross sections are obtained by multiplying the observed signal strength $\mu$ by the expected theoretical cross section of the signal MC simulated sample. The theoretical cross section for the $\PW\PGg\PGg$ and $\PZ\PGg\PGg$ signals obtained from \MGvATNLO at NLO accuracy are $18.70\pm 0.03\MCstat\pm 0.12\PDFscale\fb$ and $5.96\pm 0.01\MCstat\pm 0.06\PDFscale\fb$, respectively.
\par}

\begin{figure}[htb]
\centering
\includegraphics[width=0.49\textwidth]{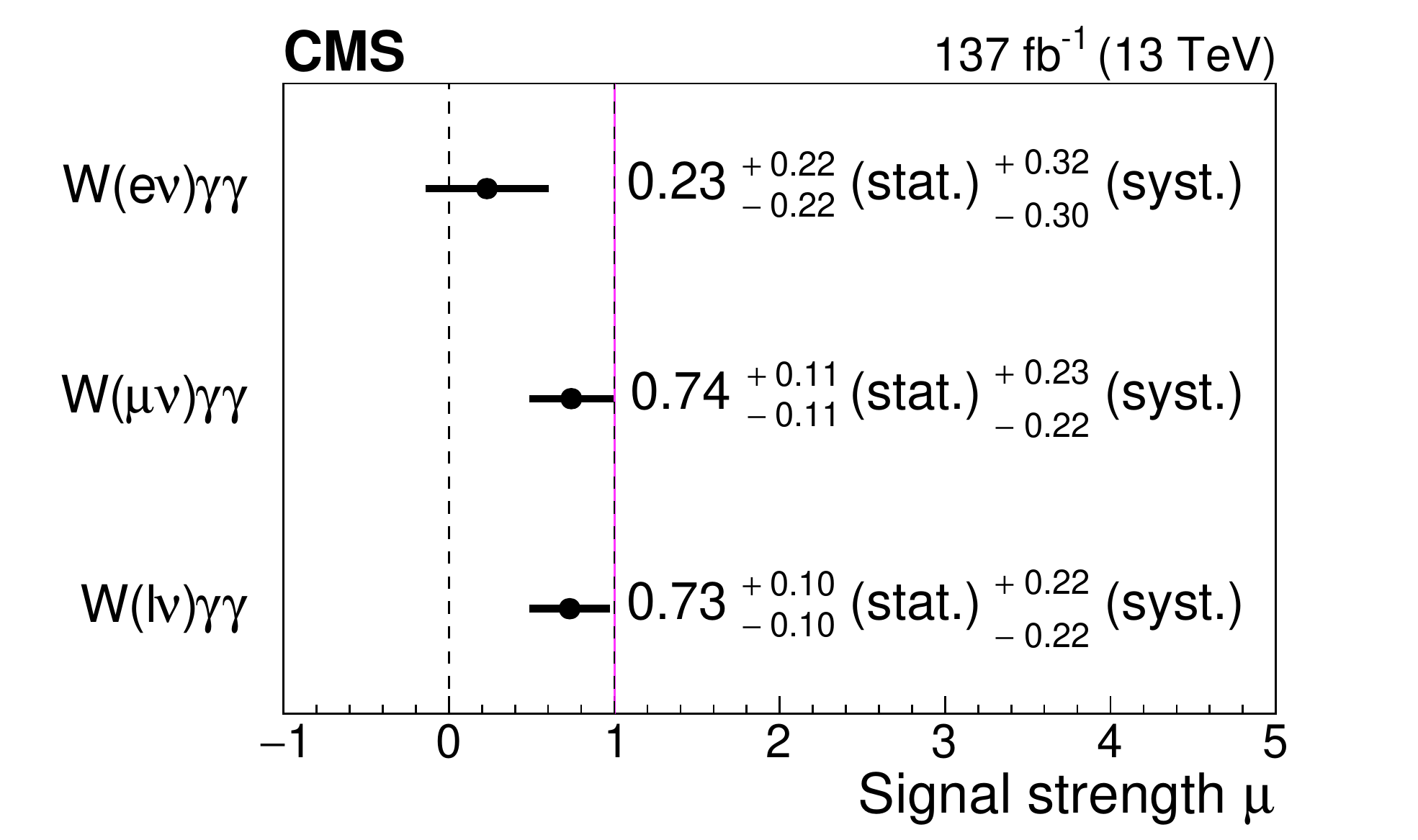}
\includegraphics[width=0.49\textwidth]{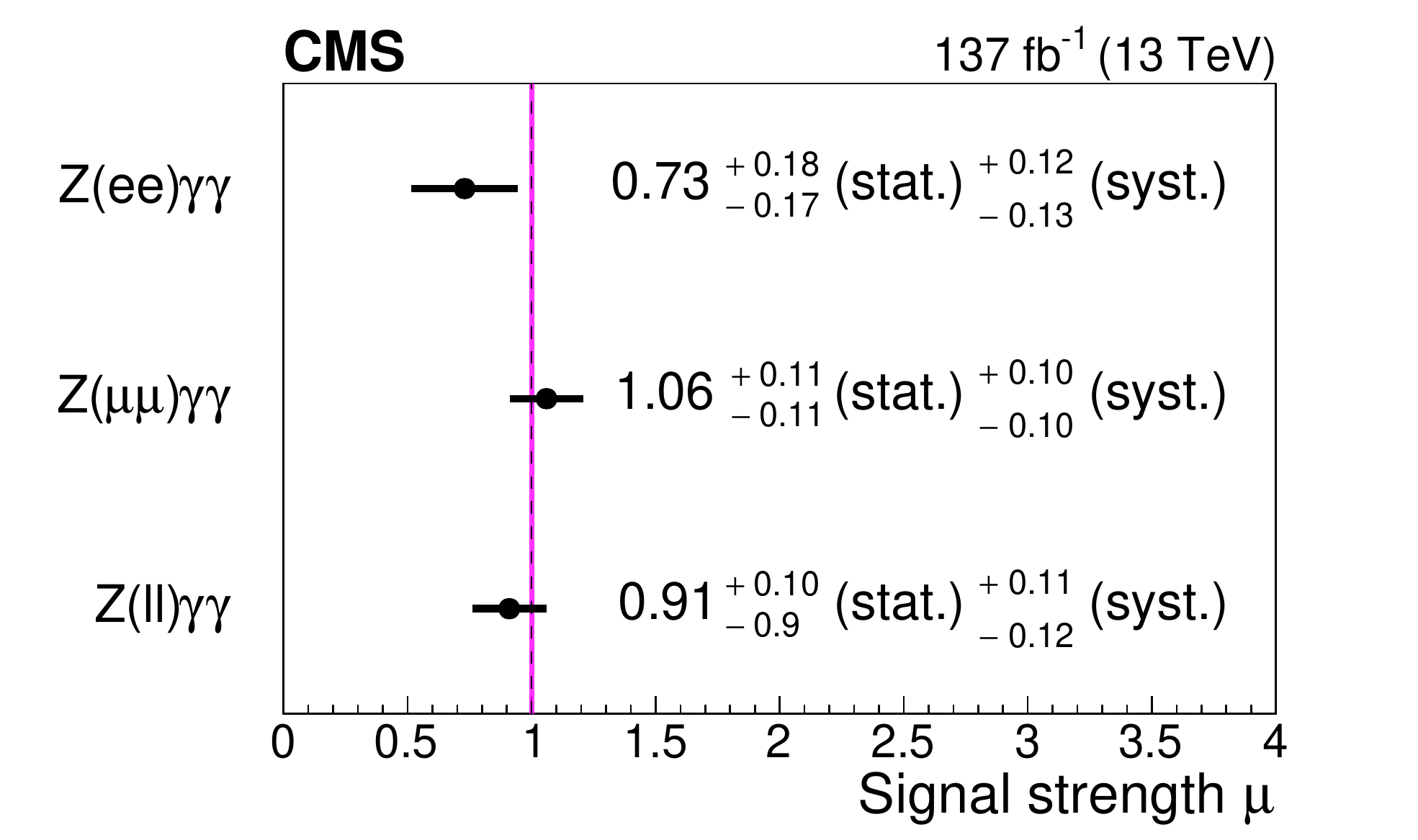}
\vspace*{\baselineskip}
\caption{Best fit values of the signal strengths for the $\PW\PGg\PGg$ (left) and $\PZ\PGg\PGg$ (right) channels. The error bars represent the total uncertainty while the magenta bands represent the theoretical uncertainty in the \MGvATNLO cross section.}
\label{f_signalstrength}
\end{figure}

In the electron channel, the best fit value for the $\PW\PGg\PGg$ signal strength is $0.23^{+0.22}_{-0.22}\stat^{+0.32}_{-0.30}\,\text{(syst)}$ and for the $\PZ\PGg\PGg$ signal strength is $0.73^{+0.18}_{-0.17}\stat^{+0.12}_{-0.13}\,\text{(syst)}$. The measured cross sections are:
\begin{linenomath}
\begin{equation*}
\begin{aligned}
  \sigma(\PW\PGg\PGg)^\text{meas}_{\Pe\PGn}&=4.4^{+4.1}_{-4.1}\stat^{+6.0}_{-5.5}\syst\pm 0.03\PDFscale\fb,\\
  \sigma(\PZ\PGg\PGg)^\text{meas}_{\Pe\Pe}&=4.35^{+1.05}_{-0.99}\stat^{+0.71}_{-0.77}\syst\pm 0.05\PDFscale\fb.\\
\end{aligned}
\end{equation*}
\end{linenomath}
In the muon channel, the best fit value for the $\PW\PGg\PGg$ signal strength is $0.74^{+0.11}_{-0.11}\stat^{+0.23}_{-0.22}\syst$ and for the $\PZ\PGg\PGg$ signal strength is $1.06^{+0.11}_{-0.11}\stat^{+0.10}_{-0.10}\syst$. The measured cross sections are:
\begin{linenomath}
\begin{equation*}
\begin{aligned}
  \sigma(\PW\PGg\PGg)^\text{meas}_{\PGm\PGn}&=13.8^{+2.1}_{-2.1}\stat^{+4.3}_{-4.2}\syst\pm 0.08\PDFscale\fb,\\
  \sigma(\PZ\PGg\PGg)^\text{meas}_{\PGm\PGm}&=6.29^{+0.67}_{-0.64}\stat^{+0.57}_{-0.58}\syst\pm 0.07\PDFscale\fb.\\
\end{aligned}
\end{equation*}
\end{linenomath}
{\tolerance=800
The results of the fit for the electron and muon channels separately are compatible within two sigmas. In the combined electron and muon channel, the best fit value for the $\PW\PGg\PGg$ signal strength is $0.73^{+0.10}_{-0.10}\stat^{+0.22}_{-0.22}\syst$ and for the $\PZ\PGg\PGg$ signal strength is $0.91^{+0.10}_{-0.09}\stat^{+0.11}_{-0.12}\syst$. The measured cross sections are:
\par}
\begin{linenomath}
\begin{equation*}
\begin{aligned}
  \sigma(\PW\PGg\PGg)^\text{meas}&=13.6^{+1.9}_{-1.9}\stat^{+4.0}_{-4.0}\syst\pm 0.08\PDFscale\fb,\\
  \sigma(\PZ\PGg\PGg)^\text{meas}&=5.41^{+0.58}_{-0.55}\stat^{+0.64}_{-0.70}\syst\pm 0.06\PDFscale\fb.\\
\end{aligned}
\end{equation*}
\end{linenomath}
The sensitivity for the $\PW\PGg\PGg$ cross section measurement is dominated by the muon channel.
The measured signal strengths are summarised in Fig.~\ref{f_signalstrength}.

The significance of the cross section measurement for both the $\PW\PGg\PGg$ and $\PZ\PGg\PGg$ channels is quantified using the background-only hypothesis under the asymptotic approximation~\cite{Cowan:2010js}. The observed (expected) significance for the $\PW\PGg\PGg$ signal is $0.6\,(2.7)\,\sigma$ in the electron channel and $3.0\,(4.3)\,\sigma$ in the muon channel; for the $\PZ\PGg\PGg$ is $3.4\,(5.0)\,\sigma$ in the electron channel and $5.4\,(5.1)\,\sigma$ in the muon channel; the combined significance for the $\PW\PGg\PGg$ is $3.1\,(4.5)\,\sigma$ and for the $\PZ\PGg\PGg$ is $4.8\,(5.8)\,\sigma$.

\section{Limits on anomalous quartic gauge couplings}\label{sec:limits}

Studies of the anomalous gauge couplings can be performed in the EFT framework~\cite{degrande201321} by expanding the SM Lagrangian to include terms with dimension higher than four. In particular, both the $\PW\PGg\PGg$ and $\PZ\PGg\PGg$ processes are sensitive to the presence of \mbox{dimension-6} and \mbox{dimension-8} operators~\cite{Eboli:2006wa}. Because of the available statistics in the $\PV\PGg\PGg$ channel, the sensitivity to \mbox{dimension-6} operators is expected to be lower than the one in the diboson production. The contribution of each operator is proportional to a coupling constant $f$ and to the inverse of the energy scale $\Lambda$ at which the new phenomena appear.

In the generation of the anomalous couplings samples, a calculation using 10 (8) different \mbox{dimension-8} operators was performed for the $\PW\PGg\PGg$ ($\PZ\PGg\PGg$) process. The operators can be divided into two subsets: the $\mathcal{L}_{\mathrm{M}0}\mdash\mathcal{L}_{\mathrm{M}7}$ ones, that contain both the $\mathrm{SU(2)}_L$ and $\mathrm{U(1)}_Y$ field strengths and the covariant derivative of the Higgs doublet, and the $\mathcal{L}_{\mathrm{T}0}\mdash\mathcal{L}_{\mathrm{T}9}$ ones, that contain only the two field strengths. In particular, the $\PW\PGg\PGg$ channel is especially sensitive to the M2, M3, T0, T1, T2, T5, T6, and T7 operators, whereas the $\PZ\PGg\PGg$ channel is especially sensitive to the T0, T1, T2, T5, T6, T7, T8, and T9 operators.

{\tolerance=800
The distribution of the \pt of the diphoton system (shown in Fig.~\ref{f_vgg_diphopt}) is used to constrain the aQGC parameters under the hypothesis of absence of anomalies in triple gauge couplings. The contribution of aQGCs is enhanced at high values of the \pt of the diphoton system.
The distribution of the aQGCs as a function of the couplings themselves has a quadratic behaviour, and hence a parabolic fit is implemented to interpolate between the different values obtained via the parameter scan. The fitting procedure is performed bin-by-bin to exploit the shape of the distribution to set the limits and include the different systematic uncertainties. To further increase the sensitivity, electron and muon channels are combined. 
Each operator coefficient is scanned independently with all other operators set to zero. The extraction of the 95\% confidence level upper and lower limits on the aQGCs is performed by exploiting the procedure described in Ref.~\cite{Khachatryan:2014jba}. The expected and measured limits for both the $\PW\PGg\PGg$ and $\PZ\PGg\PGg$ processes are presented in Table~\ref{t_agc}.
\par}

In particular, the intervals computed for the $f_{\mathrm{T}5}$ and $f_{\mathrm{T}6}$ parameters are the most constraining ones in the $\PW\PGg\PGg$ channel and are comparable to the most stringent results obtained by the $\PW\PGg\mathrm{jj}$~\cite{Sirunyan:2020azs} analysis of the CMS Collaboration at 13\TeV. The intervals computed for the $f_{\mathrm{T}0}$ and $f_{\mathrm{T}5}$ parameters in the $\PW\PGg\PGg$ channel are more stringent than the ones obtained by the $\PZ\PGg\PGg$~\cite{Aad:2016sau} and $\PZ\PGg\mathrm{jj}$~\cite{Aaboud:2017pds} analyses of ATLAS at 8\TeV. For the $\PZ\PGg\PGg$ channel, the most stringent interval is the one computed for the $f_{\mathrm{T}9}$ parameter, which is competitive with the results obtained by the $\PZ\PGg\mathrm{jj}$~\cite{CMS:2020ioi} and $\PZ\PZ\mathrm{jj}$~\cite{CMS:2020fqz} analyses of CMS at 13\TeV. The intervals computed for the $f_{\mathrm{T}8}$ and $f_{\mathrm{T}9}$ parameters in the $\PZ\PGg\PGg$ channel are more stringent than the ones obtained by the $\PZ\PGg\PGg$ and $\PZ\PGg\mathrm{jj}$ analyses of ATLAS at 8\TeV.

\begin{table}
\topcaption{Expected and observed 95\% confidence level intervals for the different anomalous couplings in both the $\PW\PGg\PGg$ and $\PZ\PGg\PGg$ channels.}
\centering
\begin{tabular}{ccccc}
& \multicolumn{2}{c}{$\PW\PGg\PGg\,(\text{Te\hspace{-.08em}V}^{-4})$} & \multicolumn{2}{c}{$\PZ\PGg\PGg\,(\text{Te\hspace{-.08em}V}^{-4})$} \\
Parameter & Expected & Observed & Expected & Observed \\
\hline
$f_{\mathrm{M}2}/\Lambda^4$ & $[-57.3,\,57.1]$ & $[-39.9,\,39.5]$ & \NA & \NA\\
$f_{\mathrm{M}3}/\Lambda^4$ & $[-91.8,\,92.6]$ & $[-63.8,\,65.0]$ & \NA & \NA\\
$f_{\mathrm{T}0}/\Lambda^4$ & $[-1.86,\,1.86]$ & $[-1.30,\,1.30]$ & $[-4.86,\,4.66]$ & $[-5.70,\,5.46]$\\
$f_{\mathrm{T}1}/\Lambda^4$ & $[-2.38,\,2.38]$ & $[-1.70,\,1.66]$ & $[-4.86,\,4.66]$ & $[-5.70,\,5.46]$\\
$f_{\mathrm{T}2}/\Lambda^4$ & $[-5.16,\,5.16]$ & $[-3.64,\,3.64]$ & $[-9.72,\,9.32]$ & $[-11.4,\,10.9]$\\
$f_{\mathrm{T}5}/\Lambda^4$ & $[-0.76,\,0.84]$ & $[-0.52,\,0.60]$ & $[-2.44,\,2.52]$ & $[-2.92,\,2.92]$\\
$f_{\mathrm{T}6}/\Lambda^4$ & $[-0.92,\,1.00]$ & $[-0.60,\,0.68]$ & $[-3.24,\,3.24]$ & $[-3.80,\,3.88]$\\
$f_{\mathrm{T}7}/\Lambda^4$ & $[-1.64,\,1.72]$ & $[-1.16,\,1.16]$ & $[-6.68,\,6.60]$ & $[-7.88,\,7.72]$\\
$f_{\mathrm{T}8}/\Lambda^4$ & \NA & \NA & $[-0.90,\,0.94]$ & $[-1.06,\,1.10]$ \\
$f_{\mathrm{T}9}/\Lambda^4$ & \NA & \NA & $[-1.54,\,1.54]$ & $[-1.82,\,1.82]$ \\
\end{tabular}
\label{t_agc}
\end{table}

\section{Summary}

The cross sections for both the $\PW\PGg\PGg$ and $\PZ\PGg\PGg$ processes are measured in proton-proton collisions by the CMS experiment at a centre-of-mass energy of 13\TeV corresponding to an integrated luminosity of 137\fbinv.

The cross sections are measured in a fiducial region where simulated signal events are selected at generator level in the $\PW\PGg\PGg$ channel by requiring exactly one electron or muon with transverse momentum $\pt>30\GeV$ and at least two photons, each with $\pt>20\GeV$. Events are selected in the $\PZ\PGg\PGg$ channel by requiring two oppositely charged electrons or muons, at least one of them with $\pt>30\GeV$, and at least two photons, each with $\pt>20\GeV$. All leptons and photons are required to have pseudorapidity $\abs{\eta}<2.5$. Additionally, the invariant mass of the dilepton system is required to exceed $m_{\ell\ell}>55\GeV$.

The measured cross sections are $13.6^{+1.9}_{-1.9}\stat^{+4.0}_{-4.0}\syst\pm 0.08\PDFscale\fb$ for the $\PW\PGg\PGg$ channel and $5.41^{+0.58}_{-0.55}\stat^{+0.64}_{-0.70}\syst\pm 0.06\PDFscale\fb$ for the $\PZ\PGg\PGg$ channel. These results are in agreement with the theoretical cross sections computed at next-to-leading order. The corresponding signal significances are 3.1 and 4.8 standard deviations. Limits on anomalous quartic gauge couplings are set using both channels.

\begin{acknowledgments}
  We congratulate our colleagues in the CERN accelerator departments for the excellent performance of the LHC and thank the technical and administrative staffs at CERN and at other CMS institutes for their contributions to the success of the CMS effort. In addition, we gratefully acknowledge the computing centres and personnel of the Worldwide LHC Computing Grid and other centres for delivering so effectively the computing infrastructure essential to our analyses. Finally, we acknowledge the enduring support for the construction and operation of the LHC, the CMS detector, and the supporting computing infrastructure provided by the following funding agencies: BMBWF and FWF (Austria); FNRS and FWO (Belgium); CNPq, CAPES, FAPERJ, FAPERGS, and FAPESP (Brazil); MES (Bulgaria); CERN; CAS, MoST, and NSFC (China); MINCIENCIAS (Colombia); MSES and CSF (Croatia); RIF (Cyprus); SENESCYT (Ecuador); MoER, ERC PUT and ERDF (Estonia); Academy of Finland, MEC, and HIP (Finland); CEA and CNRS/IN2P3 (France); BMBF, DFG, and HGF (Germany); GSRT (Greece); NKFIA (Hungary); DAE and DST (India); IPM (Iran); SFI (Ireland); INFN (Italy); MSIP and NRF (Republic of Korea); MES (Latvia); LAS (Lithuania); MOE and UM (Malaysia); BUAP, CINVESTAV, CONACYT, LNS, SEP, and UASLP-FAI (Mexico); MOS (Montenegro); MBIE (New Zealand); PAEC (Pakistan); MSHE and NSC (Poland); FCT (Portugal); JINR (Dubna); MON, RosAtom, RAS, RFBR, and NRC KI (Russia); MESTD (Serbia); SEIDI, CPAN, PCTI, and FEDER (Spain); MOSTR (Sri Lanka); Swiss Funding Agencies (Switzerland); MST (Taipei); ThEPCenter, IPST, STAR, and NSTDA (Thailand); TUBITAK and TAEK (Turkey); NASU (Ukraine); STFC (United Kingdom); DOE and NSF (USA).
  
  \hyphenation{Rachada-pisek} Individuals have received support from the Marie-Curie programme and the European Research Council and Horizon 2020 Grant, contract Nos.\ 675440, 724704, 752730, 765710 and 824093 (European Union); the Leventis Foundation; the Alfred P.\ Sloan Foundation; the Alexander von Humboldt Foundation; the Belgian Federal Science Policy Office; the Fonds pour la Formation \`a la Recherche dans l'Industrie et dans l'Agriculture (FRIA-Belgium); the Agentschap voor Innovatie door Wetenschap en Technologie (IWT-Belgium); the F.R.S.-FNRS and FWO (Belgium) under the ``Excellence of Science -- EOS" -- be.h project n.\ 30820817; the Beijing Municipal Science \& Technology Commission, No. Z191100007219010; the Ministry of Education, Youth and Sports (MEYS) of the Czech Republic; the Deutsche Forschungsgemeinschaft (DFG), under Germany's Excellence Strategy -- EXC 2121 ``Quantum Universe" -- 390833306, and under project number 400140256 - GRK2497; the Lend\"ulet (``Momentum") Programme and the J\'anos Bolyai Research Scholarship of the Hungarian Academy of Sciences, the New National Excellence Program \'UNKP, the NKFIA research grants 123842, 123959, 124845, 124850, 125105, 128713, 128786, and 129058 (Hungary); the Council of Science and Industrial Research, India; the Latvian Council of Science; the Ministry of Science and Higher Education and the National Science Center, contracts Opus 2014/15/B/ST2/03998 and 2015/19/B/ST2/02861 (Poland); the National Priorities Research Program by Qatar National Research Fund; the Ministry of Science and Higher Education, project no. 0723-2020-0041 (Russia); the Programa Estatal de Fomento de la Investigaci{\'o}n Cient{\'i}fica y T{\'e}cnica de Excelencia Mar\'{\i}a de Maeztu, grant MDM-2015-0509 and the Programa Severo Ochoa del Principado de Asturias; the Thalis and Aristeia programmes cofinanced by EU-ESF and the Greek NSRF; the Rachadapisek Sompot Fund for Postdoctoral Fellowship, Chulalongkorn University and the Chulalongkorn Academic into Its 2nd Century Project Advancement Project (Thailand); the Kavli Foundation; the Nvidia Corporation; the SuperMicro Corporation; the Welch Foundation, contract C-1845; and the Weston Havens Foundation (USA).
\end{acknowledgments}

\bibliography{auto_generated}
\cleardoublepage \appendix\section{The CMS Collaboration \label{app:collab}}\begin{sloppypar}\hyphenpenalty=5000\widowpenalty=500\clubpenalty=5000\vskip\cmsinstskip
\textbf{Yerevan Physics Institute, Yerevan, Armenia}\\*[0pt]
A.~Tumasyan
\vskip\cmsinstskip
\textbf{Institut f\"{u}r Hochenergiephysik, Wien, Austria}\\*[0pt]
W.~Adam, J.W.~Andrejkovic, T.~Bergauer, S.~Chatterjee, M.~Dragicevic, A.~Escalante~Del~Valle, R.~Fr\"{u}hwirth\cmsAuthorMark{1}, M.~Jeitler\cmsAuthorMark{1}, N.~Krammer, L.~Lechner, D.~Liko, I.~Mikulec, P.~Paulitsch, F.M.~Pitters, J.~Schieck\cmsAuthorMark{1}, R.~Sch\"{o}fbeck, M.~Spanring, S.~Templ, W.~Waltenberger, C.-E.~Wulz\cmsAuthorMark{1}
% ...
\vskip\cmsinstskip
% ...
1:  Also at TU Wien, Wien, Austria\\
% ...
\end{sloppypar}
\end{document}